\documentclass{appolb}
\usepackage{graphicx}

\begin{document}
\title{R\&D for a Dedicated Fast Timing Layer in the CMS Endcap Upgrade
\thanks{Presented at Workshop on PicosecondPhoton Sensors for Physics and Medical Applications}%
}
\author{Sebastian White
\address{Rockefeller University}
\\
}
\maketitle
\begin{abstract}
The PhaseII Upgrades of CMS are being planned for the High Luminosity LHC (HL-LHC) era when the mean number of 
interactions per beam crossing (``in-time pileup") is expected to reach $ \sim140-200$. The potential backgrounds arising from 
mis-associated jets and photon showers, for example, during event reconstruction could be reduced if physics objects are tagged with an
``event time". This tag is fully complementary to the ``event vertex" which is already commonly used to reduce mis-reconstruction.
Since the tracking vertex resolution is typically $\sim10^{-3}(\sim\frac{50 \mu m}{4.8cm}$ of the rms vertex distribution, whereas only
$\sim10^{-1} $(i.e. 20 vs.170 picoseconds (psec)) is demonstrated for timing, it is often assumed that only photon (i.e. EM calorimeter or
shower-max) timing is of interest. We show that the optimal solution will likely be a single timing layer which measures both charged particle and
photon time (a pre-shower layer).

\end{abstract}
  
\section{Introduction}
We are used to information being time-synchronized, as in a video at a given number of frames per second. In some ways the human brain works that way, since the thalamo-cortical rhythm synchronizes our sensory inputs into ``events"- associating a particular sound with a visual experience, for example.

	This synchronization is fundamental to the way that the LHC experiments have been designed.
	
	 We start from the fact that, for a given beam current, the best way to maximize the luminosity is to concentrate the beam in a small number of packets. This maximizes the number of protons that a given circulating proton will encounter at the experiment's collision point.	
	 
	 In 2012 (and possibly in 2015 also) the beam structure had a frequency of 20 MHz. This is a lower frequency than the accelerator was designed to deliver. Apart from concerns about safety due to higher beam current (corresponding to 400 MJoules at design), the primary reason for choosing the 20 MHz structure is that at 40 MHz there are complicated, hard to model, parasitic effects, which have long been a concern for operations planning\cite{cloud}.
	
	The packet structure of the beam has severe consequences for our physics, since it now leads to significant backgrounds due to pileup of several events in one frame. Today there are typically 20-30 events/frame. In the next years this will rise to 50 events and is expected to eventually reach 140-200 events/frame. Under the present conditions the background is already too severe to perform several physics studies (but not, up to now, preventing the ``flagshipÓ ones). 
The long-term strategy for the future of High Energy Physics, adopted over the past year by most national and international advisory groups (ie P5), gives the highest priority to exploiting the full potential of the LHC. Specifically these recommendations imply delivering an order of magnitude more integrated intensity than the accelerator or the experiments were designed to achieve.
	
	Clearly there is concern that the increased pileup in event frames will become so severe that the capabilities of the experiment will be reduced. These background considerations could lead us to operate at lower intensity (to preserve data quality) than projected by the advisory groups. 
	
	Our group has a history of doing R\&D to mitigate pileup through the use of time information\cite{CHEF}. Currently the time information recorded by ATLAS and CMS is used primarily to identify the packet that a collision occurred in. Resolving in-time pileup will require an order of magnitude improvement in time resolution($\sim 20$ psec).
	
\section{Event time}
	The use of ``event time" for pileup mitigation is very similar to the, more common, use of event vertex. In the case of ``event vertex" background from pileup is suppressed by measuring all vertex positions along the z-axis (beam direction) and then eliminating physics objects which are not consistent with originating from the vertex of interest. The primary workhorse is, of course, the tracker.
\begin{figure}[htb]
\centerline{
\includegraphics[width=12.5cm]{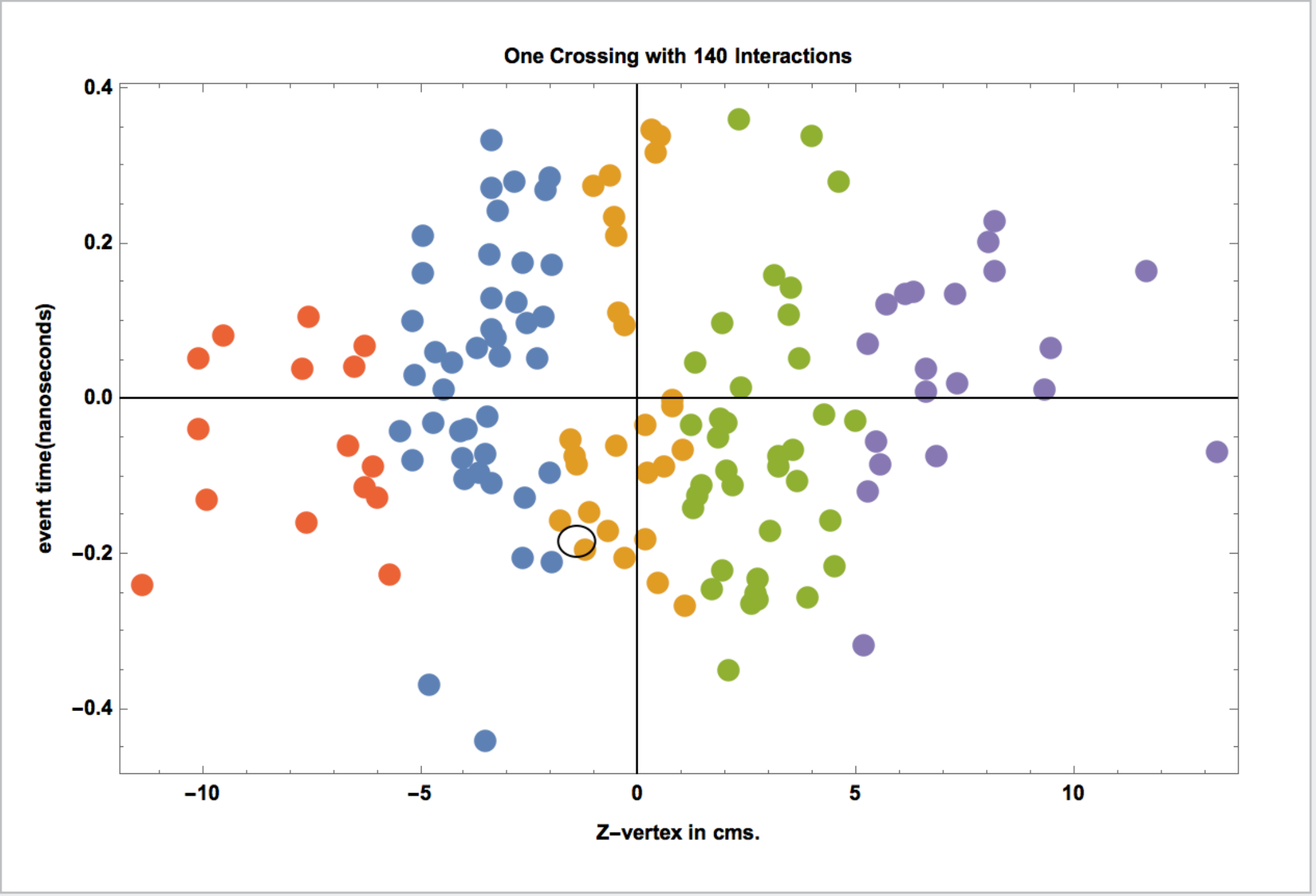}}
\caption{Simulation of the space(z-vertex) and time distribution of interactions within a single bunch crossing in CMS at a pileup of 140 events- using LHC design book for crossing angle, emittance, etc.
Typically events are distributed with an rms-in time- of 170 picoseconds, independent of vertex position.}
\label{Fig:F2H}
\end{figure}
	In the case of pileup mitigation via ``event time" the time information of physics objects of interest are also compared to an event vertex, whose time is determined from other particles in the event.
A representative frame under HL-LHC conditions is shown in Fig.1.
	Physics studies in CMS have focused on the use of timing to associate jets and electromagnetic showers with other objects in an event. It is tempting, therefore, to think of the timing device as a specialized enhancement of a calorimeter, which provides only the time of these objects. This point of view overlooks the obvious need for a device that is the workhorse, analogous to the tracker-in the case of ``event vertex" measurement, providing the ``event time" with which to correlate objects found in a calorimeter.
	
	There are actually several reasons to favor a generalized timing layer, emphasizing also the ability to measure charged track time. Aside from the need for a device, which in any case, must be
flexible enough to capture event time for vertices of interest, there is also the lack of precedent for calorimeters with $\leq \sim 100$ picosecond resolution.

	One such calorimeter ran in the low luminosity phase of ATLAS\cite{ZDC}. A large system ($\sim 16,000$ channels), based on shashlik technology, was also built and operated in the PHENIX experiment at RHIC\cite{PHENIX} and the $\sim 100 $ picosecond electromagnetic shower resolution, demonstrated in the test beam, has been used for particle identification via time of flight. A recent discussion of jet timing  performance of LHC calorimeters can be found in the CHEF2013 proceedings\cite{CHEF}. A time resolution of 200-300 picoseconds has been demonstrated by both ATLAS and CMS using 2012 data.

\subsection{Timing Layer}
	We therefore consider a dedicated timing layer in the following discussion. A layer which is primarily sensitive to charged particles has been introduced for CMS physics performance simulations
of the upgraded end cap. As a baseline this timing layer is located on the front face of the end cap EM calorimeter, with coverage extending from 1.6 $\leq\mid\eta\mid\leq$ 2.6  and a pixel size of 8x8$mm^2$. The pixel size was chosen to be fine enough to limit efficiency loss due to multiple hits/pixel at the highest $\eta$ and the full HL-LHC luminosity, calculated using FLUKA.

	This baseline can now be used for physics simulation and extended to larger $\eta$, where it could provide an important tool complementary to the tracking. There is also the open question of emphasis-ie.
\begin{itemize}
\item{a timing layer at, or near, the front of the EM calorimeter- providing most of the vertex timing info through charged hits.}
\item{a timing layer deep in the EM calorimeter- providing high efficiency for EM showers but degraded charged particle timing}
\item{possibly a single timing layer at $\sim2X_0$ could, instead,  satisfy both requirements.}
\end{itemize}

	CMS physics performance simulations over the coming year of the above options will likely guide the decision on whether to include a timing layer and how to do so.
	
	In the remainder of this article we discuss R$\&$D on detectors capable of carrying out the demanding timing measurement in the end cap, with particle fluxes of $\sim 10^7cm^{-2}$.
	
\subsection{Timing Detector Technology}

	Not surprisingly, a survey of current and planned HEP experiments quickly shows that there are no existing detectors that simultaneously meet the time precision and rate capability requirement of the CMS end cap at HL-LHC.
	
	The continuing TOF R$\&$D related to the ALICE TOF system\cite{Crispin} is now achieving the required time resolution in beam and cosmic ray tests but it does not yet have the rate capability. 
	
	A demonstration of $<10$ picosecond charged particle timing in 2006\cite{nagoya} by a Nagoya group has also been very influential. Their technique, employing a thin quartz radiator producing Cerenkov light proximity focussed on a micro-channel plate PMT(MCP-PMT), has often been copied in test beam demonstrations but never with as good performance.
	
	There are, however, a number of challenges to realizing this kind of performance in the HL-LHC application. Many of these challenges are being addressed in a long term vision of the LAPPD collaboration, which has been responsible for this conference series \cite{LAPPD}). However, in the past, this effort has been targeted at lower rate applications.

\section{New Technology}
	Our R$\&$D has, instead, focussed on developing a continuous timing layer which meets both time jitter and rate requirements, developing new detector technologies not yet in the market.
	
\subsection{Silicon sensor}
	Most of our work has focused on fast timing with silicon sensors. Starting in 2008 we started working with RMD VP for APD Research, Dick Farrell. We are developing an option, starting from their commercial deep-depleted APDs- primarily marketed as photosensors for the PET imaging community. Early measurements using $\beta$-sources had demonstrated sub-nanosecond rise time and large ($G_{APD}\sim 500$) internal gain, making this attractive for charged particle timing.
	
\begin{figure}[htb]
\begin{center}
 \includegraphics[width=0.49\textwidth]{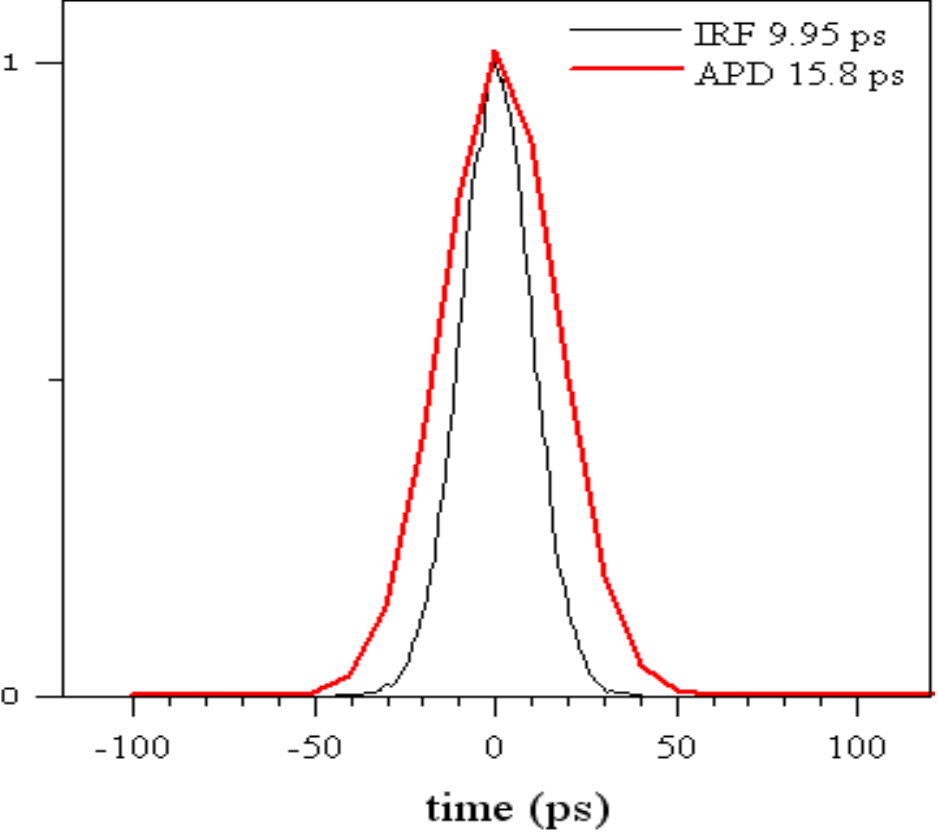}
  \includegraphics[width=0.49\textwidth]{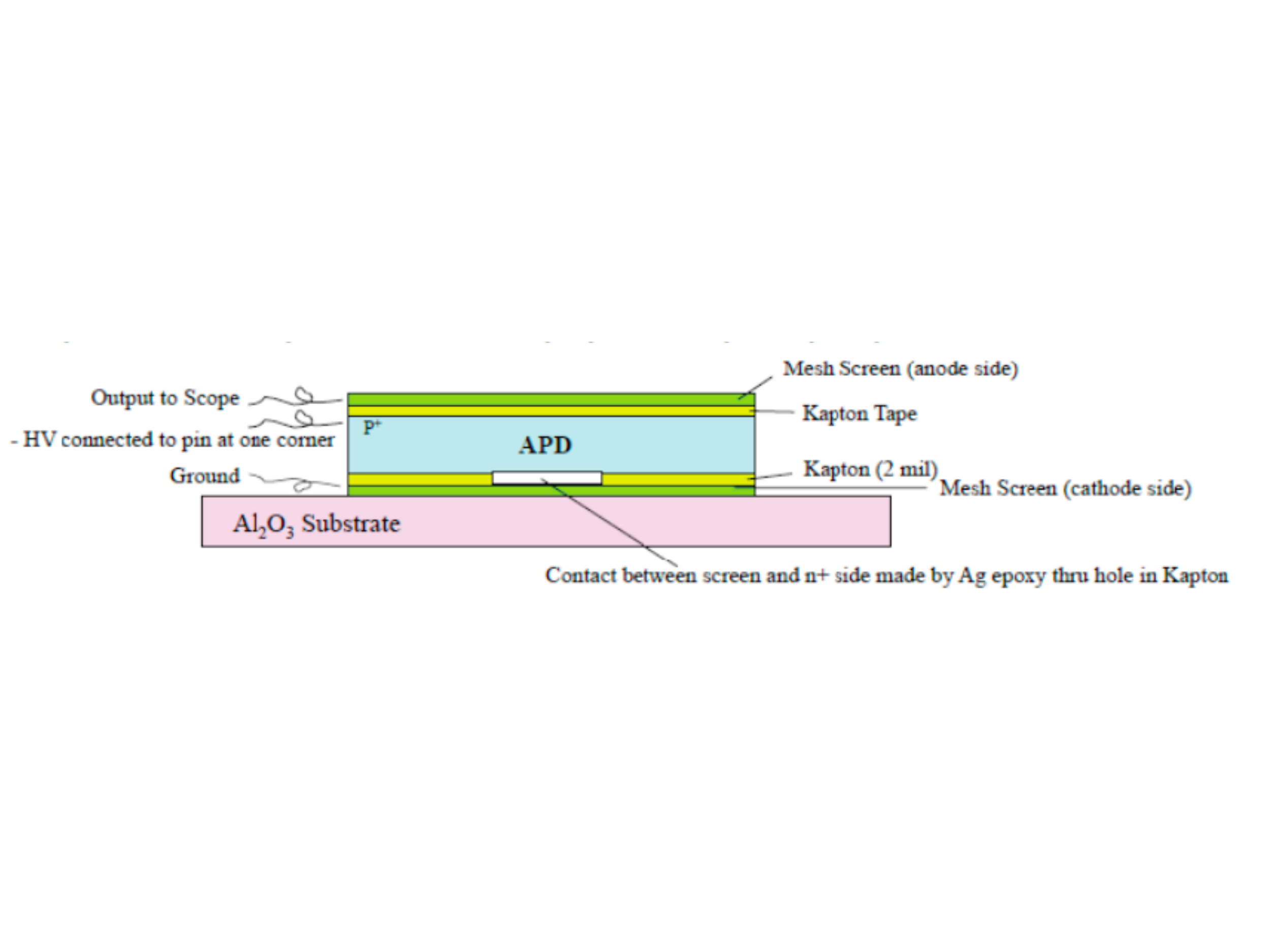}
  \caption{ APD time jitter in a 2x2$mm^2$ pixel using 980 nm femtosecond laser with spot size of $300\mu$m. and $<n_{pe}>\sim6000$, a useful model for MIP signals( {\it left}).
        In order to preserve the low time jitter over a large area APD our R$\&$D  has focused on metallization and readout  of the induced pulse on a ``MicroMegas" mesh({\it right}).   }
  \label{fig2}
 \end{center}
\end{figure}

	Early RMD radiation damage measurements and an analysis, based on CMS scaling laws for radiation damage in APDs\cite{design}, showed that these detectors would, likely, meet the rate and dose requirements for use in the CMS end cap upgrade. Specifically, up to now there has been no evidence for loss of APD gain, but the predicted increase in leakage current (due to displacement damage) is observed as well as a more rapid degradation of Quantum Efficiency (QE is irrelevant for our application as a MIP detector). However, demonstration data at higher doses are now a priority for our project. 
	
\begin{figure}[htb]
\begin{center}
 \includegraphics[width=0.49\textwidth]{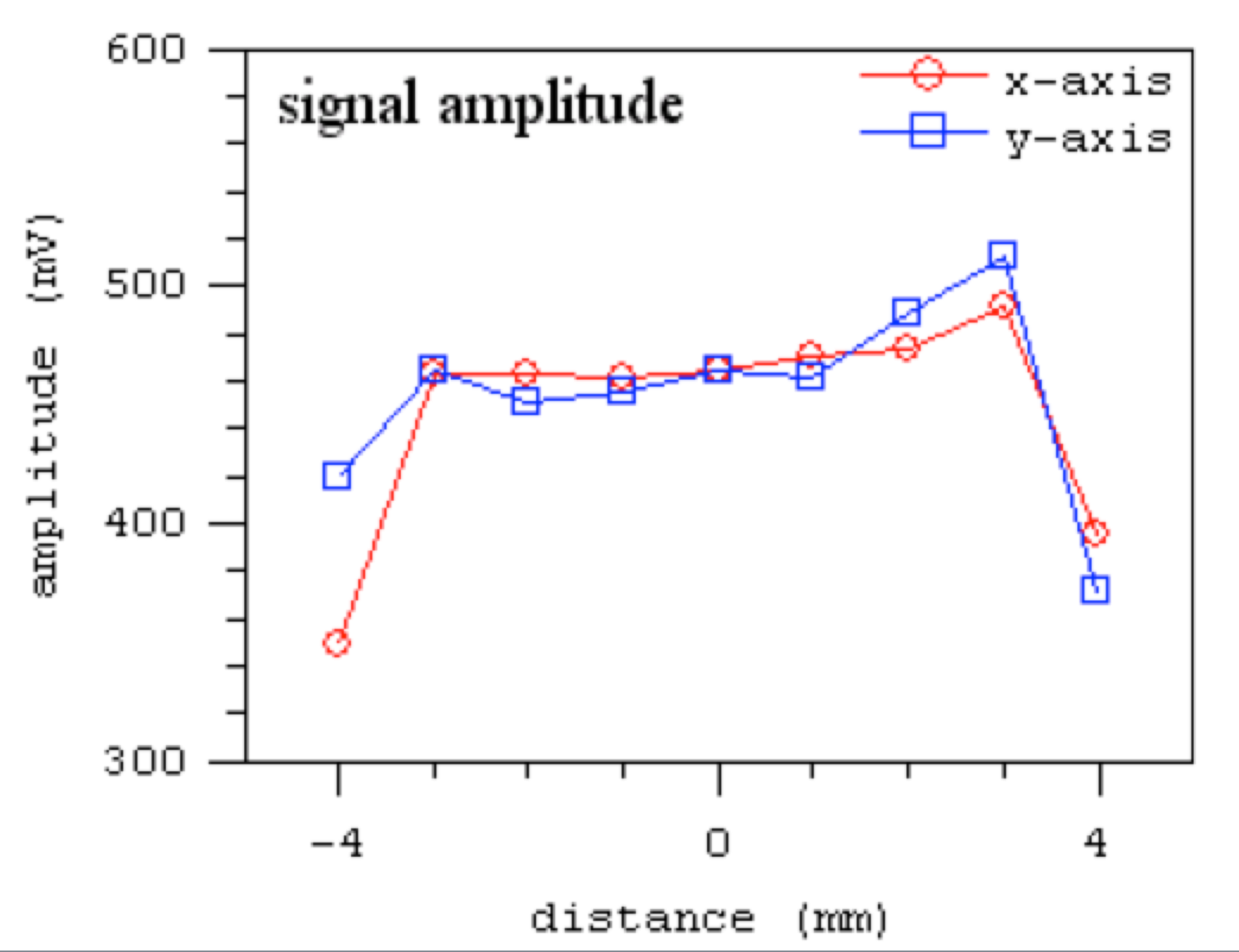}
  \includegraphics[width=0.49\textwidth]{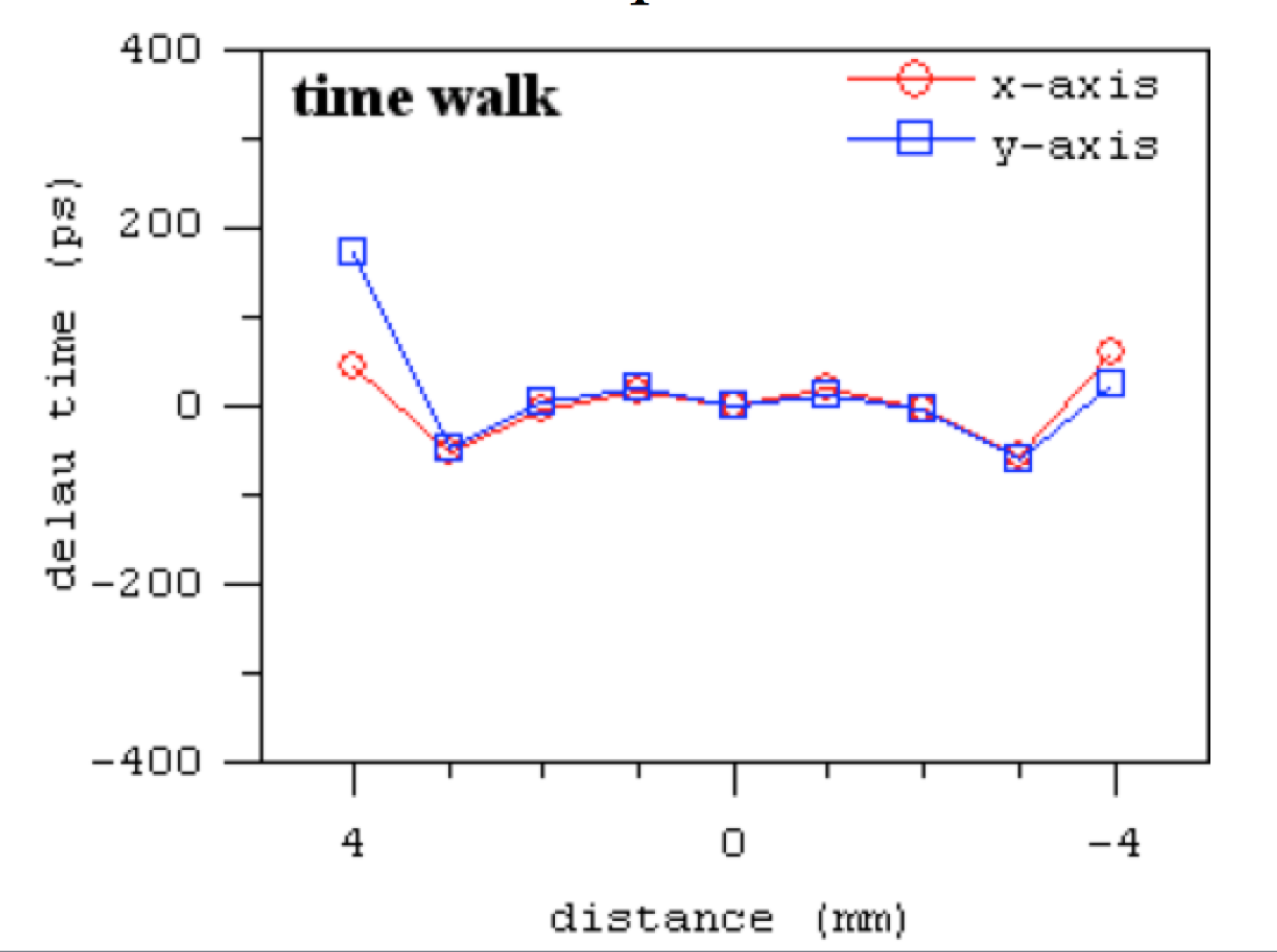}
  \caption{ Representative uniformity measurements of deep-depleted APD modified  with a mesh construction shown in Fig.2.
     ({\it left}) Amplitude variation along a horizontal scan(x) and vertical (y) .
         ({\it right}) Signal time of arrival measured as above.
       }
  \label{fig2}
 \end{center}
\end{figure}

	As shown in Fig. 2, small timing jitter can be obtained over a limited area of the APD, before modification. A useful tool for studying uniformity of response and time jitter is an IR femtosecond laser. At a wavelength of 980nm the absorption length is larger than the 40$\mu$m effective depletion thickness of these devices. A pulse intensity yielding $\sim$4000 e-h pairs in the silicon replicates the signal response to a MIP. The effect of Landau/Vavilov fluctuations, specific to MIPs is discussed in Ref.\cite{CHEF}.
	
	A representative response map for the structure shown in Fig.2 is shown in Fig. 3. The development of a larger pixel size sensor, appropriate for the CMS application, increased the effective detector capacitance, $C_D$, and circuit modeling predicts features which initially degraded the timing performance, as demonstrated in test beams at PSI and DESY.
\begin{itemize}
\item{for a $C_D$ of 50-60 pF the rise time is degraded from 700 psec to 2.0 nsec and the peak pulse is reduced by a factor of 5 when using commercial 50$\Omega$ input voltage amplifiers. This is 
being addressed by our development of a high bandwidth transimpedance amplifier based on Si-Ge technology\cite{ACES}.}
\item{depending on the internal series resistance of the APD, this larger $C_D$ could also limit timing response. The technique of reading the induced signal on the MicroMegas mesh appears to have eliminated this effect.}
\end{itemize}

	Recent progress on this technology, during the past year has included a couple rounds of prototyping of the new amplifiers, which are expected to be used in test beams at CERN or Fermilab in the coming months. We are also working with RMD on several aspects of packaging and integration with the front-end electronics. We are also in discussions with RMD concerning large scale production models, based on a revised approach where the sensor design is focused on MIP detection ab-initio.

\begin{figure}[htb]
\begin{center}
 \includegraphics[width=0.49\textwidth]{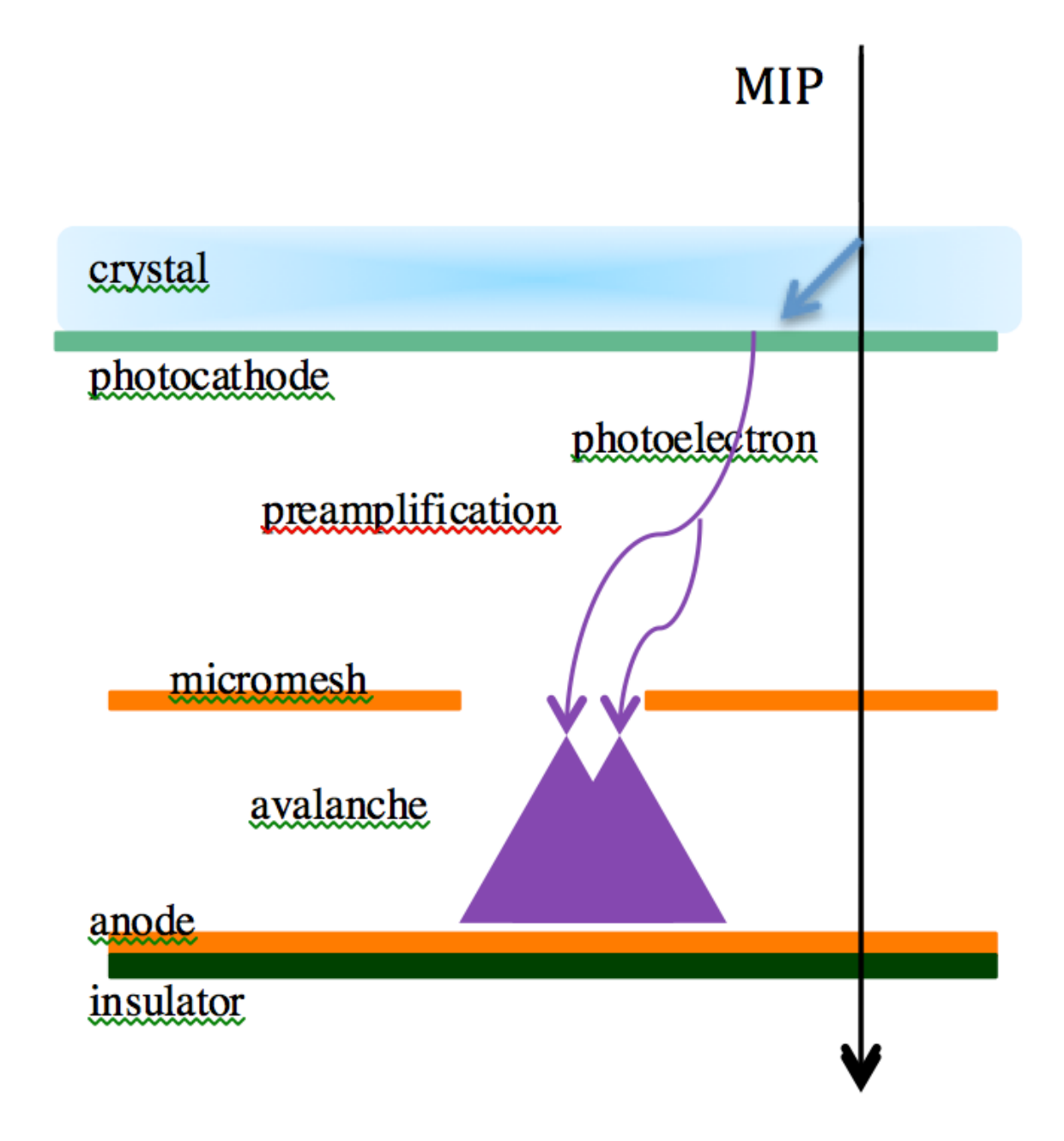}
  \includegraphics[width=0.49\textwidth]{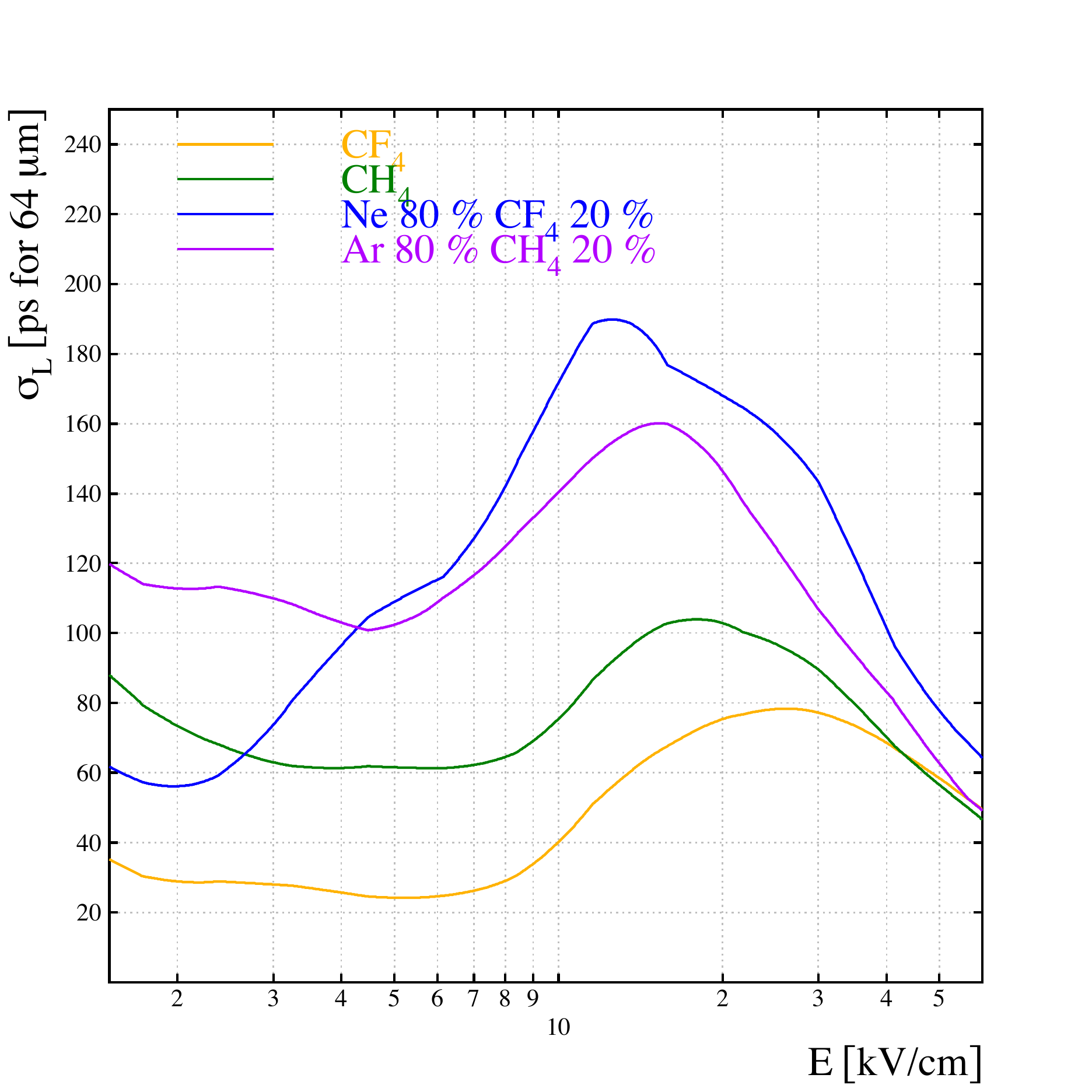}
  \caption{ Principle of Fast Gas PMT. Cerenkov photons ($\times q.e.\sim40$ photoelectrons) produced in the window produce photoelectrons,
  either in a transparent photocathode(pictured {\it left}) or a reflective one.
         ({\it right}) The diffusion-dominated time jitter can be as low as $\sim30$ picoseconds per photoelectron in a 64 micron pre-amplification gap (calculation by Rob Veenhof).
       }
  \label{fig2}
 \end{center}
\end{figure}

\subsection{MicroMegas}
	As a hedge against concerns about production costs and radiation hardness- particularly if CMS physics modeling presents a case for extended coverage (beyond $\eta=2.6$), we\cite{RD51} started detailed simulation of a Micro Pattern Gas Detector capable of delivering MIP timing at the level of $\sim20$ psec. 
	
	The principle, shown in Fig. 4, is to make an effective replacement for the MCP-PMT principle employed by the Nagoya group\cite{nagoya} for the detection of Cernekov photons- using, instead, a ``Gas PMT" principle. 
	
	Neither the time spread of the Cerenkov photons nor the diffusion in the $<100\mu$m preamplification gap in Fig. 4 would result in a MIP time jitter as large as 20 psec. We are currently building a test chamber (actually 2 different ones) for validation of this design at a pulsed UV laser facility (Saclay Laser-matter Interaction Center) in late September.
	
	If succesful, we will then construct a MIP timing detector to evaluate in a test beam. To address CMS specific requirements we will evaluate photocathode alternatives with lifetime suited to HL-LHC intensities. Another aspect of this development concerns MicroMegas mesh alternatives that satisfy the signal recovery for high rates as well as the $C_D$ issues.
	
\section{Acknowledgement}

	We are grateful for helpful discussions with Mitch Newcomer, Fritz Caspers, Crispin Williams, Erich Griesmayer, Rob Veenhof and Anatoly Ronzhin. We thank Roman Zuyeuski for frequent help with DAQ software.
I would like to thank Christophe Royon for organizing this meeting and US-CMS for continued support of this R$\&$D.

\end{document}